\documentclass[pre,epsfig,twocolumn,showpacs,preprintnumbers,amssymb]{revtex4}
\usepackage{bm,graphicx,amsmath}

\begin{document}

\title{Diffusive transport of light in two-dimensional disordered packing of
disks: Analytical approach to transport-mean-free path}

\author{Zeinab Sadjadi}
\affiliation{Institute for Advanced Studies in Basic Sciences
(IASBS), P. O. Box 45195-1159, Zanjan 45195, Iran}

\author{MirFaez Miri }
\email{miri@iasbs.ac.ir} \affiliation{Institute for Advanced
Studies in Basic Sciences (IASBS), P. O. Box 45195-1159, Zanjan
45195, Iran}

\author{M. Reza Shaebani}
\affiliation{Institute for Advanced Studies in Basic Sciences
(IASBS), P. O. Box 45195-1159, Zanjan 45195, Iran}

\author{ Sareh Nakhaee}
\affiliation{Institute for Advanced Studies in Basic Sciences
(IASBS), P. O. Box 45195-1159, Zanjan 45195, Iran}

\begin{abstract}
We study photon diffusion in a two-dimensional random packing of
monodisperse disks as a {simple} model of granular media and wet
foams.
%To simplify our model
We assume that the intensity reflectance of disks is a constant
$r$. We present an {\it analytic} expression for the
transport-mean-free path $l^{\ast}$ in terms of the velocity of
light in the disks and host medium, radius $R$ and packing
fraction of the disks, and the intensity reflectance. For the
glass beads immersed in the air or water, we estimate
transport-mean-free paths about half the experimental ones. For
the air bubbles immersed in the water, $l^{\ast}/R$ is a {\it
linear} function of $ 1/\varepsilon$, where $ \varepsilon$ is the
liquid volume fraction of the model wet foam. This throws new
light on the empirical law of Vera et. al [Applied Optics
\textbf{40}, 4210 (2001)], and promotes more realistic models.
\end{abstract}

\pacs{05.40.Fb, 42.25.Dd, 82.70.Rr, 81.05.Rm}

%42.25.Dd    Wave propagation in random media
%82.70.Rr    Aerosols and foams
%81.05.Rm    Porous materials, granular materials
%42.68.Ay    Propagation, transmission, attenuation, and radiative transfer

\maketitle

\section{Introduction}

There is a good reason to study wave propagation in turbid or
random media: Multiply scattered waves can {\it probe} temporal
changes in physical systems\ \cite{maret1,Sneider,sheng}. Thus,
light transport through fog\ \cite{fog}, milky liquids, nematic
liquid crystals\ \cite{DWSnematic}, granular media\
\cite{granular1, granular1b, granular1c, granular2}, foams\
\cite{Durianold, vera, Gittings2004, gopal, hoo1, hoo2, hoo3}, and
human tissue\ \cite{Yodh95}; propagation of elastic waves in the
Earth's crust\ \cite{ludocel,ludocel2}; acoustic waves in the
fluidized or sedimenting suspensions\ \cite{page}; etc., have
attracted much attention.

In a turbid medium, light undergoes many scattering events before
leaving the sample, and the transport of light energy is
diffusive\ \cite{sheng}. Therefore, the photon can be considered
as a random walker. The transport-mean-free path $l^{\ast}$, over
which the photon direction becomes randomized, depends on the
structural details of the opaque medium. Experimental techniques
like diffuse-transmission spectroscopy (DTS)\ \cite{kaplan} and
diffusing-wave spectroscopy (DWS)\ \cite{DWS} can be used to
measure $l^{\ast}$. In DTS, the average fraction $T$ of incident
light transmitted through a slab of thickness $L$ is measured. The
transport-mean-free path is then deduced from $T \propto
l^{\ast}/L$. Utilizing the temporal intensity fluctuations in the
speckle field of the multiply scattered light, DWS determines
$l^{\ast} $ and the mean-squared displacement of the scattering
sites due to time evolution, thermal motion, or flow.
%Lemiueux PRE 1998 Np4498
%fog

A plethora of light-scattering experiments show that light
transport reaches its diffusive limit in granular media\
\cite{granular1, granular1b, granular1c, granular2} and foams\
\cite{Durianold, vera, Gittings2004, gopal, hoo1, hoo2, hoo3},
which means that photons perform a random walk. However, the
mechanisms underlying this random walk are not elucidated. A {wet}
foam is composed of spherical gas bubbles dispersed in liquid. A
relatively {dry} foam consists of polyhedral cells separated by
thin liquid films. Three of them meet in the so-called Plateau
borders which then define tetrahedral vertices\ \cite{Weaire1999}.
In their studies of foams with the liquid volume fraction
$\varepsilon$ in the range $0.008<\varepsilon<0.3$, Vera,
Saint-Jalmes, and Durian\ \cite{vera} observed the empirical law
\begin{equation}
 l^{\ast} \approx 2 R(\frac{0.14}{\varepsilon}+1.5 ), \label{empi}
\end{equation}
where $R$ is the average bubble radius. Recent studies of
scattering from Plateau borders\ \cite{vera, sun1}, vertices\
\cite{Skipetrov02}, and films\ \cite{miriA, miriB, miriC, miriD,
miriE, miriE2}; or transport effects such as total internal
reflection of photons inside the Plateau
borders\cite{Gittings2004, miriF}, have not yet clarified the
empirical law of Vera et al.. For granular media, systematic
measurements of the transport-mean-free path $l^{\ast}$ as a
function of the refractive indices of grains and the host medium
(air, water,...), grain size, and packing fraction, have not been
performed. Menon et al.\ \cite{granular1} determined $l^{\ast}
\approx 15 R $ for glass spheres of radius $R=47.5 \mathrm{ \mu m}
$ dispersed in air. For glass beads dispersed in water, Leutz et
al.\ \cite{granular1c} found $l^{\ast} \approx 14R-16R $ for $80
\mathrm{ \mu m} \leqslant R\leqslant 200 \mathrm{ \mu m} $. Their
samples had a packing fraction $ \phi \approx 0.64 $. Crassous\
\cite{granular2} performed numerical simulations to find
$l^{\ast}$ as a function of refractive indices of the grain and
host medium, but only for packing fraction $ \phi \approx 0.64 $.

It is instructive to consider {\it simple} or even toy models of
granular media and wet foams, which allow an {\it analytic} access
to the transport-mean-free path $l^{\ast}$. Apparently, such
models pave the way for deeper understanding of fascinating DWS
experiments. In this paper, we consider two-dimensional packing of
monosize disks. The disks are much larger than the wavelength of
light, thus one can employ ray optics to follow a light beam or
photon as it is reflected by the disks with a probability $r$
called the intensity reflectance. We assume that the intensity
reflectance is constant, and the velocity of light inside and
outside the disks are ${c}/{n_{in}}$  and ${c}/{n_{out}}$,
respectively. We show that the photon's random walk based on the
above rules is a persistent random walk\ \cite{miriA, kehr, w1}.
Writing a master equation to describe the photon transport, we
find in Sec.~\ref{AA} the transport-mean-free path as
\begin{equation}
l^{\ast}=   \frac{\pi R}{4}
 \frac{(\frac{3}{2r}-1) \Big(\frac{r}{1-r}+(\frac{1-\phi}{\phi})^2\Big)}{(\frac{\phi}{ n_{in}}+
\frac{1-\phi}{n_{out}})
(n_{in}\frac{r}{1-r}+n_{out}\frac{1-\phi}{\phi})}
 , \label{mainresult}
\end{equation}
where $\phi$ and $R$ denote the packing fraction and radius of
disks, respectively. We further study our model by numerical
simulation of the photon's random walk. We observe the overall
agreement between our numerical and analytical estimates of the
transport-mean-free path.

For glass beads immersed in air or water, we find
transport-mean-free paths about half the experimental ones\
\cite{granular1, granular1c}. For the air bubbles immersed in the
water, we use Eq.~(\ref{mainresult}) to derive $l^{\ast}$ as a
function of the liquid volume fraction $\varepsilon=1-\phi$. $r
\approx 0.20 $ is estimated as a weighted average of Fresnel's
intensity reflectance. We find that in the range $0.08 <
\varepsilon<0.15 $, our {analytical} result agrees well with the
relation $ l^{\ast} \approx R({0.11}/{\varepsilon}+2.37)$. In
other words, we find that $l^{\ast}/R$ is a {\it linear} function
of $1/\varepsilon$. Using the hybrid lattice gas model for
two-dimensional foams and Fresnel's intensity reflectance, Sun and
Hutzler performed numerical simulation of photon transport and
found $ l^{\ast} \approx R({0.26}/{\varepsilon}+4.90 )$\
\cite{sun1}. Quite remarkably, our {\it analytic} estimate of the
transport-mean-free path throws new light on the empirical law of
Vera et al. and the numerical simulation of Sun et al..

Our article is organized as follows. In Section~\ref{model} we
introduce the two-dimensional packing of disks as a simple model
for a granular medium or a wet foam. Photon transport in a random
packing of disks using constant intensity reflectance is discussed
in Sec.~\ref{mm}. Discussions, conclusions, and an outlook are
presented in Sec.~\ref{dis}.

\section{Model} \label{model}

As a {\it simple} model for a two-dimensional disordered granular
medium, wet foam, and bubbly liquid, we choose the random packing
of circular disks. All non-overlapping disks have the same radius
$R$, and cover a fraction $\phi$ of the plane. To address the
photon transport in such medium, we have made the following
assumptions: (i) Disks or grains, are much larger than the
wavelength of light, thus one can employ ray optics to follow a
light beam or photon as it is reflected by the disks with a
probability $r$ called the intensity reflectance. (ii) $r$ is a
constant, with no dependence on the incidence angle. (iii)
Although disks of refractive index $n_{in}$ are immersed in a
medium of refractive index $n_{out}$, the incident and the
transmitted rays have the same direction. In other words, we
assume that the angle of refraction equals the angle of incidence.
(iv) The velocity of light inside and outside the disks are
${c}/{n_{in}}$ and ${c}/{n_{out}}$, respectively.

Our first assumption is inspired by the experiments\
\cite{granular1, granular1b, granular1c, granular2}. Our second
and third assumptions do not agree with Fresnel's formulas and
Snell's law, respectively. Consequently, our model does not
consider total internal reflection of rays. However, we
deliberately adopt a step-by-step approach to photon transport in
granular media, and will consider more realistic models later.

As already mentioned, we model single photon paths in a packing of
disks as a random walk with rules motivated by ray optics,
{\i.e.}, an incoming light beam is reflected from a disk surface
with a probability $r$ or it traverses the disk surface with a
probability $t=1-r$. This naturally leads to a persistent random
walk of the photons\ \cite{miriA}, where the walker remembers its
direction from the previous step\ \cite{kehr, w1}. Persistent
random walks are employed in biological problems\ \cite{furth},
turbulent diffusion \cite{tay}, polymers\ \cite{flory}, Landauer
diffusion coefficient for a one-dimensional solid\ \cite{55}, and
in general transport mechanisms\ \cite{t0,t0b}. More recent
applications are reviewed in\ \cite{w2}. In the following section,
we adopt the approach of\ \cite{miriC, t0} to study persistent
random walk of the photons in a granular medium.

\section{Photon transport in a two-dimensional packing of disks}\label{mm}
\subsection{Analytical treatment} \label{AA}

\begin{figure}
\includegraphics[width=0.8\columnwidth]{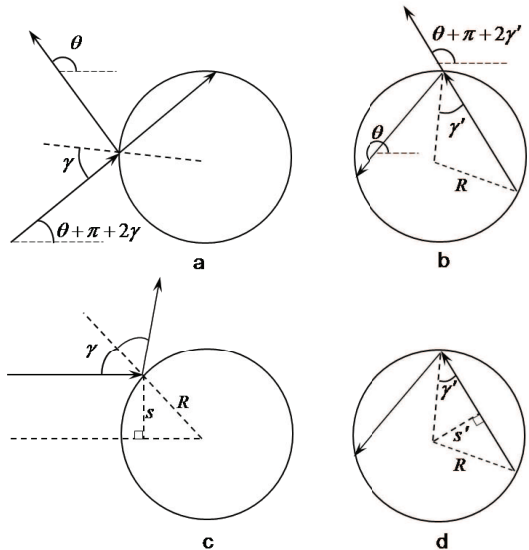}
\caption{(a) Path of a photon moving in the host medium and
hitting a disk with an incidence angle $\gamma$. (b) Path of a
photon moving in a disk and hitting its surface with an incidence
angle $\gamma'$. The step length inside the disk is $ 2 R \cos
\gamma'$, where $R$ is radius of the disk. (c) A photon impinging
on a disk with an impact parameter $s$. Note $s=R \sin\gamma$. (d)
The distance $s'$ is related to the incidence angle $\gamma' $ by
$s'=R \sin\gamma'$.
 }
\label{f1}
\end{figure}

The photon random walk in a packing of disks consists of steps
inside and outside the disks. We denote the average length of
steps inside and outside the grains by $\bar{L}_{in}$ and
$\bar{L}_{out}$, respectively. We characterize each step by an
angle relative to the $x$-axis. As Fig.~\ref{f1}(a) demonstrates,
on hitting a disk with an incidence angle $\gamma $, a photon
moving in the host medium along the direction $\theta+\pi+2\gamma
$ will be either reflected to the direction $ \theta$, or enters
the disk. The probability distribution of the random variable
$\gamma $ ($ 0 < \gamma<\pi/2 $) is $F(\gamma)=\cos \gamma $, see
Appendix\ \ref{app1}. Similarly, a photon moving in a disk along
the direction $\theta+\pi+2\gamma' $ and hitting its surface with
an angle $\gamma'$, will be either reflected to the direction $
\theta$ or enters the host medium, see Fig.~\ref{f1}(b). Quite
remarkably, the probability distribution of the incidence angles
$\gamma $ and $\gamma' $ are the same, see Appendix\ \ref{app1}.
We can therefore conclude that diffusion of the photons inside and
outside the grains are not inherently different. As will be shown
in the following, we write a master equation to describe the
photon diffusion inside (outside) the grains, utilizing the step
length $\bar{L}_{in}$ ($\bar{L}_{out}$) and velocity
${c}/{n_{in}}$ (${c}/{n_{out}}$), and extract the diffusion
constant $D_{in}$ ($D_{out}$). According to the two-state model of
Lennard-Jones\ \cite{w1}, the diffusion constant of photons in the
granular medium is
\begin{equation}
{D}_{m}=f_{in}D_{in}+f_{out}D_{out}, \label{lenard}
\end{equation}
where $f_{in}$ ($f_{out}=1-f_{in}$) is the fraction of time that
the photons spend inside (outside) the disks.

We introduce the probability $P_{n}(x,y|\theta) dx dy$ that the
photon after its $n$th step of length $ \bar{L}$ along the
direction $\theta$, arrives in the area $dx dy $ at position
$\mathbf{x}=(x,y)$. Then the following master equation expresses
the evolution of $P_{n}(x,y|\theta)$:
\begin{widetext}
\begin{equation}
 P_{n+1}(x,y|\theta)= \frac{1}{2}\enspace r
 \int_{-\frac{\pi}{2}}^{\frac{\pi}{2}}  P_{n}(x-\bar{L}\cos\theta, y-\bar{L}\sin\theta |
 \theta+\pi+2\gamma)~
   F(\gamma) d\gamma + t \, P_{n}(x- \bar{L} \cos\theta, y - \bar{L}
    \sin \theta | \theta) . \label{master}
\end{equation}
\end{widetext}
The first term on the right-hand side describes the reflection of
the photon with a probability $r$. The photon which has arrived at
position $(x-\bar{L}\cos\theta,y-\bar{L}\sin\theta )$ along the
direction $\theta+\pi+2\gamma $, changes its direction by an angle
$\pi+2\gamma $ according to the probability function $F(\gamma)$\
\cite{comment}. The second term describes the transmission with a
probability $t=1-r$. The photon performs a ballistic motion with
step length $\bar{L}$ along direction $\theta$ from position
$(x-\bar{L}\cos\theta,y-\bar{L}\sin\theta )$ to $(x,y)$.

The diffusion constant follows from the evaluation of the second
moment of $P_{n}(x,y|\theta)$ with respect to the spatial
coordinates $x$ and $y$. The probability distribution as an exact
solution of the master equation (\ref{master}) is hard to obtain.
However, there is a more direct method for the evaluation of the
moments which employs the characteristic function $\mathbf{P}_{n}(
\omega_x, \omega_y |m ) $ associated with $P_{n}(x,y|\theta)$\
\cite{w1}:
\begin{eqnarray}
\langle x^{k_{1}} y^{k_{2}} \rangle_n &\equiv& \int \int \int
x^{k_{1}} y^{k_{2}}  P_{n}(x,y|\theta) dx dy d\theta \nonumber \\
&=&
 \left. (-i)^{k_{1}+k_{2}} \frac{\partial^{k_{1}+k_{2}}
\mathbf{P}_{n}(\vec{\omega} |m=0 )}{\partial \omega_{x}^{k_{1}}
\partial \omega_{y}^{k_{2}}} \right|_{\vec{\omega}=\mathbf{0}} ,
\label{mean}
\end{eqnarray}
%where $ k_1$ and $ k_2$ are arbitrary nonnegative integers,
where $ k_1$ and $ k_2$ are either zero or positive
integers,
\begin{equation} \mathbf{P}_{n}( \vec{\omega} |m )
 \equiv \int_{-\pi}^{\pi} e^{i m \theta} \int \int e^{i \vec{\omega}
\cdot\mathbf{x}} P_{n}(x,y |\theta ) dx dy  d\theta
,\label{charac}
\end{equation}
and $\vec{\omega} = (\omega_x, \omega_y )$. We are interested in
the first and second moments of $P_{n}(x,y|\theta)$, thus we focus
on the Taylor expansion
\begin{eqnarray}
 \mathbf{P}_{n}( \omega, \alpha  |m ) &\approx& Q_{0,n}(\alpha|m) +
i \omega \bar{L} Q_{1,n}(\alpha|m) \nonumber \\
& &- \frac{\omega^2 \bar{L}^2 }{2} Q_{2,n}(\alpha|m)+ \dots ,
\label{taylor}
\end{eqnarray}
where  $\omega$ and $\alpha $ are the polar representation of the
vector $\vec{\omega} = (\omega_x, \omega_y  )$. From
Eqs.~(\ref{mean}) and (\ref{taylor}) it follows that
\begin{eqnarray}
  \langle x \rangle_n &=& \bar{L} Q_{1,n}(0|0), \nonumber \\
\langle y \rangle_n &=& \bar{L} Q_{1,n}( \frac{\pi}{2}|0), \nonumber \\
\langle x^2 \rangle_n &=& \bar{L}^2 Q_{2,n}(0|0), \nonumber \\
\langle y^2 \rangle_n &=& \bar{L}^2 Q_{2,n}( \frac{\pi}{2}|0) .
\label{meanmean}
\end{eqnarray}

Fourier transforming Eq.~(\ref{master}), we obtain
\begin{eqnarray}
\mathbf{P}_{n+1}( \omega, \alpha  |m )&=&
\sum_{k=-\infty}^{\infty} i^k e^{-i k \alpha} J_k(\omega \bar{L})
\mathbf{P}_{n}( \omega, \alpha |k+m )   \nonumber \\
 & &\times
[r(-1)^{m+k}\mathbf{F}(2m+2k)+t] , \label{cagefourier}
\end{eqnarray}
where  $\mathbf{F}(m)= \frac{1}{2}\int_{-{\pi}/{2}}^{{\pi}/{2}}
e^{i m \gamma} F(\gamma) d\gamma  $, and
\begin{equation}
J_k(z) = \frac{1}{2\pi i^{k}} \int_{-\pi}^{\pi} e^{iz\cos\theta}
e^{-ik\theta} d\theta
\end{equation}
is the $k$th-order Bessel function. Since we are only interested
in the Taylor coefficients $Q_{1,n}(\alpha|m)$ and
$Q_{2,n}(\alpha|m)$, we insert Eq.~(\ref{taylor}) into
Eq.~(\ref{cagefourier}). Using the Taylor expansion of the
relevant Bessel functions $J_k(z)$ ($|k| \le 2$) and $J_k(0) =
\delta_{0,k}$\ \cite{arf} and collecting all terms with the same
power in $\omega$, results in the following recursion relations
for the $Q_{i,n}(\alpha|m)$:
\begin{eqnarray}
&& Q_{0,n+1}(\alpha|m)=\big[t+ r (-1)^m \mathbf{F}(2m) \big]  Q_{0,n}(\alpha|m)  ,\nonumber \\
    \nonumber\\
&&Q_{1,n+1}(\alpha|m)= \big[t+r (-1)^m \mathbf{F}(2m) \big] Q_{1,n}(\alpha|m) \nonumber \\
&&\hspace{0.7cm} + \frac{e^{-i \alpha} }{2}\big[t+r(-1)^{m+1}\mathbf{F}(2m+2)\big] Q_{0,n}(\alpha|m+1)   \nonumber \\
& & \hspace{0.7cm}+ \frac{e^{i \alpha}}{2} \big[t+r(-1)^{m-1}\mathbf{F}(2m-2)\big]   Q_{0,n}(\alpha|m-1)  ,\nonumber\\
&& Q_{2,n+1}(\alpha|m)=\big[ t+ r (-1)^m \mathbf{F}(2m)\big]    Q_{2,n}(\alpha|m) \nonumber\\
&& \hspace{0.7cm} + e^{-i \alpha} \big[t+r(-1)^{m+1}\mathbf{F}(2m+2)\big]      Q_{1,n}(\alpha|m+1)  \nonumber\\
&& \hspace{0.7cm} +
 e^{i\alpha}\big[t+r(-1)^{m-1}\mathbf{F}(2m-2)\big] Q_{1,n}(\alpha|m-1)\nonumber \\
& &  \hspace{0.7cm} +\frac{1 }{2} \big[t+ r (-1)^m \mathbf{F}(2m) \big]    Q_{0,n}(\alpha|m)  \nonumber \\
&&  \hspace{0.7cm} + \frac{e^{-2i \alpha}}{4}  \big[ t+r(-1)^{m+2}\mathbf{F}(2m+4) \big] Q_{0,n}(\alpha|m+2) \nonumber \\
&&  \hspace{0.7cm} + \frac{e^{2i \alpha}}{4}
\big[t+r(-1)^{m-2}\mathbf{F}(2m-4) \big] Q_{0,n}(\alpha|m-2).
\nonumber \\ \label{setQQ}
\end{eqnarray}
We solve this set of coupled linear difference equations using the
method of the $z$-transform\ \cite{w1, jury}. The $z$-transform
${Q}(z)$ of a function $Q_n$ of a discrete variable $n=0, 1, 2,
...$ is defined by $ {Q}(z)=\sum_{n=0}^{\infty} Q_n z^n$. One then
derives the $z$-transform of $Q_{n+1}$ simply as $Q(z)/z -
Q_{n=0}/z$. Note the similarities of this rule with the Laplace
transform of the time derivative of a continuous function\
\cite{arf}. The $z$-transform of equations (\ref{setQQ}) leads to
a set of algebraic equations which immediately gives
\begin{widetext}
\begin{eqnarray}
Q_{0}(z,\alpha|m)&=& \frac{Q_{0,n=0}(\alpha|m)}{1-z\big[t+r(-1)^m\mathbf{F}(2m)\big]} ,\nonumber \\
    \nonumber\\
Q_{1}(z,\alpha|m)&=&
\frac{Q_{1,n=0}(\alpha|m)}{1-z\big[t+r(-1)^m\mathbf{F}(2m)\big]} +
\frac{z}{2\big(1-z\big[t+r(-1)^m\mathbf{F}(2m)\big]\big)} \nonumber\\
&\times &\bigg\{\frac{e^{-i
\alpha}\big[t+r(-1)^{m+1}\mathbf{F}(2m+2)\big]
Q_{0,n=0}(\alpha|m+1)}{1-z\big[t+r(-1)^{m+1}\mathbf{F}(2m+2)\big]}+
    \frac{e^{i \alpha}\big[t+r(-1)^{m-1}\mathbf{F}(2m-2)\big]  Q_{0,n=0}(\alpha|m-1)}{1-z\big[t+r(-1)^{m-1}\mathbf{F}(2m-2)\big]}\bigg\},\nonumber\\
    \nonumber\\
Q_{2}(z,\alpha|m)&=&
\frac{Q_{2,n=0}(\alpha|m)}{1-z\big[t+r(-1)^m\mathbf{F}(2m)\big]}+
    \frac{zQ_{0,n=0}(\alpha|m)}{2\big(1-z\big[t+r(-1)^m\mathbf{F}(2m)\big]\big)^2}+\frac{z}{1-z\big[t+r(-1)^m\mathbf{F}(2m)\big]}\nonumber \\
&\times&     \bigg\{e^{-i
\alpha}\big[t+r(-1)^{m+1}\mathbf{F}(2m+2)\big]
Q_{1}(z,\alpha|m+1)+
e^{i \alpha}\big[t+r(-1)^{m-1}\mathbf{F}(2m-2)\big] Q_{1}(z,\alpha|m-1)\nonumber \\
  &+&\frac{e^{-2i\alpha}\big[t+r(-1)^m\mathbf{F}(2m+4)\big]Q_{0,n=0}(\alpha|m+2)}{4\big(1-z\big[t+r(-1)^{m+2}\mathbf{F}(2m+4)\big]\big)}+
   \frac{e^{2i
   \alpha}\big[t+r(-1)^m\mathbf{F}(2m-4)\big]Q_{0,n=0}(\alpha|m-2)}{4\big(1-z\big[t+r(-1)^{m-2}\mathbf{F}(2m-4)\big]\big)}\bigg\},
\end{eqnarray}
\end{widetext}
where $ \mathbf{F}(m)=(1-m^2)^{-1}\cos({m\pi}/{2}) $, especially
$\mathbf{F}(0)=1 $. The above expressions contain the sum of
several terms whose inverse $z$-transform are readily accessible:
\begin{eqnarray}
1 & \leftrightarrow & \frac{1}{1-z} ,\nonumber \\
n  & \leftrightarrow & \frac{z}{(1-z)^2}, \nonumber \\
a^n & \leftrightarrow & \frac{1}{1-a z} , \nonumber \\
n a^n & \leftrightarrow & \frac{az}{(1-a z)^2}.
\end{eqnarray}
Here $a$ is an arbitrary real number whose absolute magnitude is
less than $1$.

For an arbitrary initial distribution $P_{0}(x,y|\theta)$, the
relevant function $Q_{1,n}(\alpha|0)$ contains terms which are
either constant or behave as $a^{n}$ with $|a|<1$. They are
associated with the randomization of the initial distribution of
the random walkers but are not essential for large $n$. According
to Eq.~(\ref{meanmean}),
\begin{equation}
\langle x \rangle_{n}= \langle y \rangle_{n} = 0 \label{mean1}
\end{equation}
The behavior of the mean-square displacements is associated with
$Q_{2,n}(\alpha|0)$, see Eq.~(\ref{meanmean}). We checked that for
large $n$ or in the long-time limit it is purely diffusive, i.e.,
\begin{eqnarray}
\langle x^2 \rangle_n &=& 2 D_x \tau  , \nonumber \\
\langle y^2 \rangle_n &=& 2 D_y \tau  ,
\end{eqnarray}
where we introduced the time $\tau = n \bar{L}/v$ which passes
when the random walker makes $n$ steps at a speed $v$. We extract
the diffusion constants from $Q_{2,n}(0|0)$ and $Q_{2,n}(
\frac{\pi}{2}|0)$:
\begin{equation}
D_x=D_y=\frac{1}{4}\bar{L}v(\frac{3}{2r}-1) \label{finalD}
\end{equation}

As already mentioned, we write the master equation (\ref{master})
to describe photon diffusion in the grains and in the host medium.
Equation (\ref{finalD}) immediately leads to
\begin{eqnarray}
D_{in}   &=&\frac{1}{4}\bar{L}_{in}   \frac{c}{n_{in}}
(\frac{3}{2r}-1),
\nonumber \\
D_{out}&=&\frac{1}{4}\bar{L}_{out} \frac{c}{n_{out}}
(\frac{3}{2r}-1).
\end{eqnarray}
The task is now expressing $\bar{L}_{in}$, $\bar{L}_{out}$,
$f_{in}$ and $f_{out}$ in terms of the model parameters $R$,
$\phi$, $r$, and using the two-state model of Lennard-Jones to
derive $D_m$. First, we note that $\bar{L}_{in}= <2R \cos\gamma'>
$, where $\gamma'$ is the incidence angle of photons moving in the
disk, and here $<>$ denotes averaging with respect to the
probability distribution $F'(\gamma')=\cos \gamma' $, see
Appendix\ \ref{app1} and Fig.~\ref{f1}(b). Second,
$\phi={\bar{L}_{in}}/({\bar{L}_{in}+\bar{L}_{out}})$. Hence we
find
\begin{eqnarray}
\bar{L}_{in}&=&\frac{\pi R}{2}  ,\nonumber \\
\bar{L}_{out}&=&\frac{\pi R}{2}\frac{1-\phi}{\phi}. \label{pppp}
\end{eqnarray}
The evaluation of $f_{in}$ is more exacting. Each step length
$\bar{L}_{in}$ inside a disk takes a time $
\bar{\tau}_{in}=\bar{L}_{in} n_{in}/c$. The probability of $m$
internal steps before leaving the disk is $ t r^m$. Hence the
average time that a photon spends in the disk is $ \sum_{m=0} m
\bar{\tau}_{in}  t r^m=  \bar{\tau}_{in} r/t $. The photon spends
a time $  \bar{\tau}_{out}=\bar{L}_{out} n_{out}/c $ before
reaching a disk. Hence
\begin{eqnarray}
f_{in}&=& \frac{ \bar{\tau}_{in} r/t}{ \bar{\tau}_{out}+  \bar{\tau}_{in} r/t}=\frac{n_{in}\frac{r}{1-r}}{n_{in}\frac{r}{1-r}+n_{out}\frac{1-\phi}{\phi}},\nonumber \\
f_{out}&=&\frac{ \bar{\tau}_{out} }{ \bar{\tau}_{out}+
\bar{\tau}_{in} r/t}
=\frac{n_{out}\frac{1-\phi}{\phi}}{n_{in}\frac{r}{1-r}+n_{out}\frac{1-\phi}{\phi}}.
\label{exacting}
\end{eqnarray}
Now we utilize Eq.~(\ref{lenard}) to derive the diffusion constant
of photons in the granular medium as
\begin{equation}
D_m=\frac{\pi R c}{8} \frac{ (\frac{3}{2r}-1)
\Big(\frac{r}{1-r}+(\frac{1-\phi}{\phi})^2\Big)}{n_{in}\frac{r}{1-r}+n_{out}\frac{1-\phi}{\phi}}.
\label{lkj0}
\end{equation}
In two-dimensional space, the transport-mean-free path follows
from
\begin{equation}
l^{\ast} = 2 D_m/{c}_m , \label{lkj1}
\end{equation}
where ${c}_m$ is the transport velocity of light in the medium. To
a first approximation
%where ${c}_m$ is the velocity of light in the medium. To a good
%approximation
\begin{equation}
{c}_m=\phi \frac{c}{n_{in}} +(1-\phi)\frac{c}{n_{out}}.
\label{lkj2}
\end{equation}
Note that the velocity of light in the disks (the host medium)
covering a fraction $\phi$ ($1-\phi$) of the plane is $c/n_{in}$
($c/n_{out}$). From Eqs.~(\ref{lkj0}-\ref{lkj2}), we find the
transport-mean-free path $l^{\ast}$ mentioned already in
Eq.~(\ref{mainresult}) in the introduction.

\subsection{Numerical simulations} \label{BB}

\begin{figure}
\includegraphics[width=0.5\columnwidth]{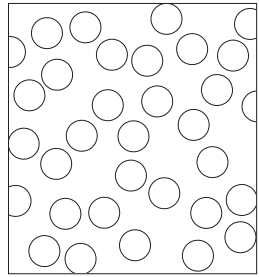}
\caption{Part of a packing of $ 10^4$ disks, covering a fraction
$\phi=0.35$ of the plane. } \label{f2a}
\end{figure}

\begin{figure}
\includegraphics[width=1.0\columnwidth]{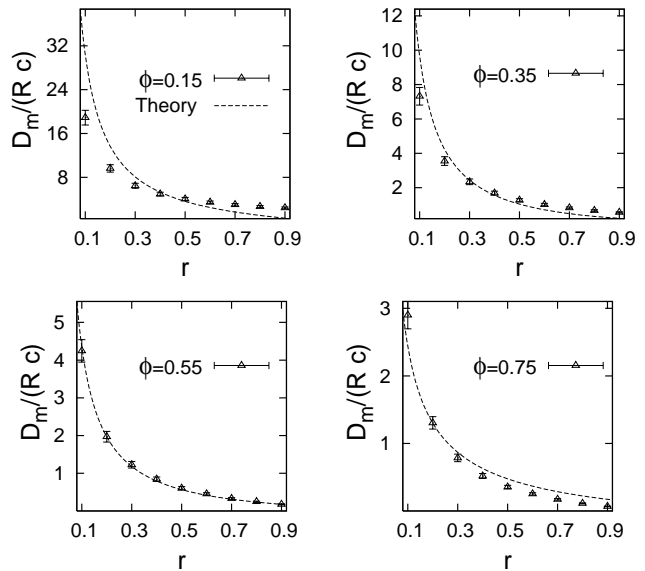}
\caption{ The diffusion constant $D_m$ (in units of the disk
radius $R$ times the velocity of light $c$) as a function of the
intensity reflectance $r$, for various packing fractions. Here
$n_{in}=1.5$ and $n_{out}=1.0$ are assumed. Monte Carlo simulation
results and $D_m(r)$ are denoted, respectively, by points and the
line.
 }
\label{f2}
\end{figure}

We presented an {\it analytic} theory to calculate the diffusion
constant of photons. Now we carefully examine this analytic result
by performing numerical simulations.

In order to generate random configurations of monodisperse disks
with a desired packing fraction, we compress a dilute system of
rigid disks into a smaller space. Simulation methods based on a
confining box generate a packing whose properties in the vicinity
of walls differ from those in the bulk. Hence we utilize the
compaction method of Ref.\ \cite{Shaebani08jcp} which combines the
contact dynamics algorithm\ \cite{Jean99,Brendel04} with the
concept of the Andersen dynamics\ \cite{Andersen80}. This combined
simulation method involves variable area of the simulation box
with periodic boundary conditions in all directions. Due to the
exclusion of side walls, the algorithm generates homogenous
packings.

We let photons perform a random walk in our packing of disks by
applying the rules introduced in Sec.~\ref{model}. For improving
the speed of our ray tracing program, we adopt the cell index
method commonly used in the molecular dynamics simulations\
\cite{allen}. The square simulation box is divided into a regular
lattice of $J \times J$ cells. We maintain a list of disks in each
of these cells. A photon moving in the cell $j$ ($ 1\leqslant j
\leqslant J^2$) probably hits the disks in the cell $j$ or its
neighbor cells. Thus it is not necessary to check collision
between the photon and {\it all} disks of the medium.

Our computer program shrinks an initial dilute sample of $10^4$
non-overlapping disks randomly distributed in a two-dimensional
simulation box. In the course of shrinking the packing, the
program saves snapshots of the grain positions if the packing
fraction $ \phi \in [0.15, 0.25,...,0.75]$, see Fig.~\ref{f2a}.
For each medium, the program takes $10^4$ photons at an initial
position, and launches them in a direction specified by angle
$\theta_0$ relative to the $x$-axis. Then it generates the
trajectory of each photon following a standard Monte Carlo
procedure and evaluates the statistics of the photon cloud at
times $\tau \in [7000,7100 ,...,9900]$ (in units of $ R /c $). As
detailed in Ref.\ \cite{miriA}, we determine the diffusion
constant $D_m$ from the temporal evolution of the average
mean-square displacement of the photons: $\langle x^2+y^2 \rangle
= 4 D_m \tau$.

\begin{figure}
\includegraphics[width=1.0\columnwidth]{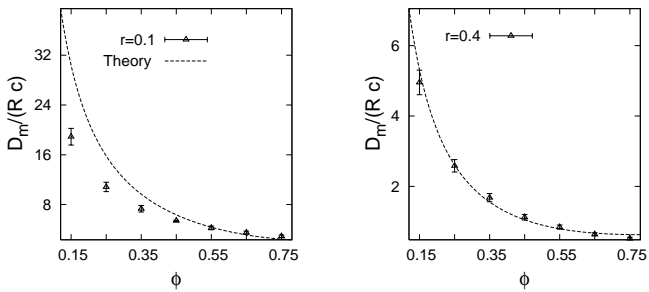}
\caption{ The diffusion constant $D_m$ (in units of the disk
radius $R$ times the velocity of light $c$) as a function of the
packing fraction $ \phi$, for $r=0.1$ and $0.4$. Here $n_{in}=1.5$
and $n_{out}=1.0$ are assumed. Monte Carlo simulation results and
$D_m( \phi)$ are denoted, respectively, by points and the line.
 }
\label{f3}
\end{figure}

For angles $ \theta_0\in [20^{\circ}, 40^{\circ}, ..., 320^{\circ}
]$, the simulation is repeated for each intensity reflectance $r
\in [0.1, 0.2,...,0.9]$. As a reasonable result, no dependence on
the starting point and the starting direction is observed. In
Fig.~\ref{f2} we plot the diffusion constant $D_m$ (in units of
the disk radius $R$ times the velocity of light $c$) as a function
of the intensity reflectance $r$, for the glass disks
($n_{in}=1.5$) immersed in the air ($n_{out}=1.0$). For this
medium, Fig.~\ref{f3} shows the diffusion constant $D_m$ (in units
of the disk radius $R$ times the velocity of light $c$) as a
function of the packing fraction $\phi$, for the intensity
reflectances $r=0.1$ and $0.4$. The errorbars reflect the the
standard deviations when we average over all diffusion constants
%$D_m(\theta)$
for different starting positions and angles. We also performed
simulations for the other examples ($n_{in}=1.34, n_{out}=1.0$),
($n_{in}=1.0, n_{out}=1.34$), $(n_{in}=1.0, n_{out}=1.5$), etc.,
but the results are not shown here. We observed the overall
agreement between the numerical results and our theoretical value
for $D_m$. Quite remarkably, Eq.~(\ref{lkj0}) involves no free
parameters, but reasonably agrees with the numerical results in a
wide range of $ \phi$, $r$, $n_{in}$, and $ n_{out}$.

\section{Discussion, Conclusions, and Outlook}\label{dis}

Diffusing-wave spectroscopy provides invaluable information about
the static and dynamic properties of granular media\
\cite{granular1, granular1b, granular1c, granular2} and foams\
\cite{Durianold, vera, Gittings2004, gopal, hoo1, hoo2, hoo3}. The
transport-mean-free path $l^{\ast}$ in terms of the microscopic
structure, however, remains to be fully elucidated. In this paper,
we consider a {\it simple} model for photon diffusion in a
two-dimensional packing of disks. Our {\it analytical} result for
$l^{\ast}$ provides new insights into the light transport, and
promotes more realistic models.

We have studied the photon's persistent random walk in a
two-dimensional packing of monodisperse disks. We employed ray
optics to follow a light beam or photon as it is reflected by the
disks. We used a constant intensity reflectance $r$. Moreover we
assumed that on hitting a disk, the incident and transmitted rays
have the same direction. To achieve a better understanding of
photon diffusion in granular medium and wet foams, we are
extending our studies by considering Fresnel's formulae for the
intensity reflectance, Snell's law of the refraction, and the
three-dimensional packing of polydisperse spheres. We are also
improving our estimate of $l^{\ast}$ by taking into account the
distribution of photons' step length, and the transport velocity
of photons\ \cite{priv}. We discuss these points in the following.

Many {\it two}-dimensional systems offer a rich and unexpected
behavior. For an experimental observation of photon diffusion in a
{two}-dimensional packing, a set of parallel fibres can be used.
Photons injected in a plane perpendicular to the axis of fibres
perform a planar diffusion. One can also drill parallel cylinders
in a host medium, and fill all the cylinders with a liquid:
$l^{\ast}$ dependence on the refractive index $n_{in}$ can be
measured. Note that a ray maintains its polarization state on
hitting a disk, and Fresnel's intensity reflectance depends on the
polarization state: Quite interesting, the transport-mean-free
paths for the transverse electric and transverse magnetic
polarizations, are different. Inspired by the rich optics of a
two-dimensional packing and following a step-by-step approach to a
real system, our attention is naturally directed to estimate
$l^{\ast}$ of a {\it three}-dimensional granular medium or wet
foam. We speculate that one can multiply Eq.~(\ref{mainresult})
with a factor of about $1$ to estimate the transport-mean-free
path for a {three}-dimensional packing. Apparently, only a
detailed study will approve or disapprove our speculation which is
based on the following observation: To understand the role of
liquid films for light transport in dry foams, we studied the two-
and three-dimensional Voronoi foams\ \cite{miriB, miriE2}. The
interesting result is that the transport-mean-free path for these
dry foams are determined by the same factor $(1-r)/r$ for a
constant intensity reflectance $r$ in spite of the difference in
{dimension}: $l^{\ast}_{\text{2D Voronoi}}(r) \approx 1.10 {R} {
(1-r)}/{r} $ and $ l^{\ast}_{\text{3D Voronoi}}(r) \approx 1.26
{R} {(1-r)}/{r}$, where $R$ denotes the average cell radius\
\cite{comment2}.

\begin{figure}
\includegraphics[width=0.8\columnwidth]{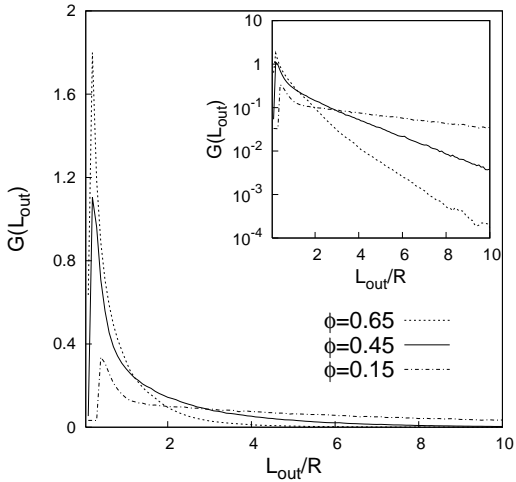}
\caption{The probability distribution $G({L}_{out})$ as a function
of ${L}_{out}/R$ for various packing fraction $\phi$. Inset: The
same plot in the logarithmic scale. Note that $G({L}_{out})$
decays exponentially.
 }
\label{gg5}
\end{figure}

Our "mean-field" theory presented in Sec.~\ref{AA} relies on the
{\it average} step lengths $\bar{L}_{in}$ and $\bar{L}_{out}$. We
note that ${L}_{in}=2 R \cos \gamma'$ and the probability
distribution of the incidence angle $\gamma'$ is $F'(\gamma')=\cos
\gamma' $, see Appendix\ \ref{app1} and Fig.~\ref{f1}(b). It
follows that the probability distribution of ${L}_{in}$ is
$G({L}_{in})={L}_{in}/(2 R \sqrt{4R^2- {L}_{in}^2})$,
$\bar{L}_{in}=\pi R/2$, and $ \overline{L^2 _{in}} = \int
{L}_{in}^2 G({L}_{in})~ d{L}_{in} =8 R^2/3$. Fig.\ \ref{gg5}
delineates the probability distribution $G({L}_{out})$ as a
function of ${L}_{out}/R$ for various packing fraction $\phi$.
After reaching its pronounced maximum, the distribution
$G({L}_{out})$ decays exponentially.
%$ \overline{L^2 _{out}}$ is different from $(\bar{L}_{out})^2$.
The pioneering work of Heiderich et al.~\cite{PLA} suggests that
the broadness of $G({L}_{in})$ and $G({L}_{out})$ affects the
value of $l^{\ast}$ by a factor about $1$.

In two-dimensional space, $l^{\ast}$ and $D_m$ are related as
$D_m= l^{\ast} {c}_m /2$, where ${c}_m$ is the transport velocity
of light. In a medium composed of dielectric spheres comparable to
the light wavelength $ \lambda$ (Mie scatterers), the transport
velocity differs by an order of magnitude from the phase velocity\
\cite{speed}. When spheres are much larger than the light
wavelength, the difference between the two velocities becomes
unimportant. However, our first approximation to ${c}_m$ [Eq.
(\ref{lkj2})] can be improved by considering the real (rather than
infinite) size parameter of disks, and the pair-correlation
function of disks.

We assumed that on hitting a disk, the incident and transmitted
rays have the same direction. Note that in the case $n_{out}>
n_{in}$, a photon moving in the host medium and hitting the disks,
experiences an average scattering angle $\int_{0}^{\pi/2} (\pi-2
\gamma) F(\gamma) d \gamma=2 $ (in radians) due to the
reflections. Taking into account inequality of the incidence and
refraction angles, and the total internal reflection of rays with
incidence angle greater than $\gamma_c= \arcsin(n_{in}/ n_{out})$,
the average scattering angle due to the transmissions is
$\int_{0}^{\gamma_c}  [ \arcsin( n_{out}/n_{in} \sin \gamma ) -
\gamma] F(\gamma) d \gamma =0.132$ (in radians), where $
n_{out}=1.34$ and $n_{in}=1$ are assumed. Reflections are more
efficient than transmissions in randomizing the direction of
photons. Thus the assumption that the incident and transmitted
rays have the same direction, leads to a plausible estimate of
$l^{\ast}$.

Fresnel's intensity reflectance depends on the incidence angle and
the polarization state of the light. However an average (constant)
intensity reflectance $r$ may describe the photon diffusion. To
estimate $r$, we first consider the case $ n_{out} > n_{in} $ and
define the critical angle $\gamma_c= \arcsin(n_{in}/ n_{out})$.
The probability distribution of the incidence angles $\gamma $ and
$\gamma'$ are $F(\gamma)=\cos \gamma$ and $F'(\gamma')=\cos
\gamma'$. We define $r_{out \rightarrow in} =\int_{0}^{\pi/2}
r_{out \rightarrow in} (\gamma) F(\gamma) d \gamma$. Here $ {out
\rightarrow in} $ indicates a ray incident from the host medium
onto the disk. The incident electric field can be decomposed into
a component parallel ($p$) to the plane of incidence, and a
component perpendicular ($s$) to this plane. To address the photon
diffusion in a {\it three}-dimensional random packing, as
reflectance $ r_{out \rightarrow in} (\gamma)$ we have taken an
average over the $p$ and $s$ polarizations: $r_{out \rightarrow
in} (\gamma)=0.5 [r_{out \rightarrow in} (\gamma,p )+ r_{out
\rightarrow in}(\gamma,s )]$. Note that for $\gamma>\gamma_c$, the
Fresnel's intensity reflectances are $1$. Similarly, we define
$r_{ in \rightarrow out} =\int_{0}^{\pi/2} 0.5 [r_{in \rightarrow
out} (\gamma',p )+ r_{in \rightarrow out}(\gamma',s )] F'(\gamma')
d \gamma'$, and then use $r=0.5(r_{out \rightarrow in}+r_{in
\rightarrow out} ) $ in Eq.~(\ref{mainresult}) to estimate
$l^{\ast}$.

In the case $ n_{out} < n_{in} $, the appropriate critical angle
is $\gamma_c= \arcsin(n_{out}/ n_{in})$. Here $0<\gamma<\pi/2$ and
$0<\gamma'<\gamma_c$. The fact that $\gamma'<\gamma_c$ ensures
that a photon moving in the disk is able to enter the host medium,
since the Fresnel's intensity reflectance is $1$ for $\gamma'>
\gamma_c$. An extension of Appendix\ \ref{app1} and numerical
simulations indicate that the probability distribution of the
incidence angles $\gamma $ and $\gamma'$ are $F(\gamma)=\cos
\gamma $ and $F'(\gamma')=\cos \gamma' / \sin \gamma_c$,
respectively. We define $r_{out \rightarrow in} =\int_{0}^{ \pi/2
} 0.5 [r_{out \rightarrow in} (\gamma,p )+ r_{out \rightarrow
in}(\gamma,s )] F(\gamma) d \gamma$ and $r_{ in \rightarrow out}
=\int_{0}^{ \gamma_c} 0.5 [r_{in \rightarrow out} (\gamma',p )+
r_{in \rightarrow out}(\gamma',s )] F'(\gamma') d \gamma'$ to
estimate $r=0.5(r_{out \rightarrow in}+r_{in \rightarrow out} )$.
In the limit $n_{in}=n_{out}$ we find $r=0$. In this limit the
photon transport is ballistic and Eq.~(\ref{mainresult}) correctly
predicts $l^{\ast}\rightarrow \infty $.

\begin{figure}
\includegraphics[width=0.7\columnwidth]{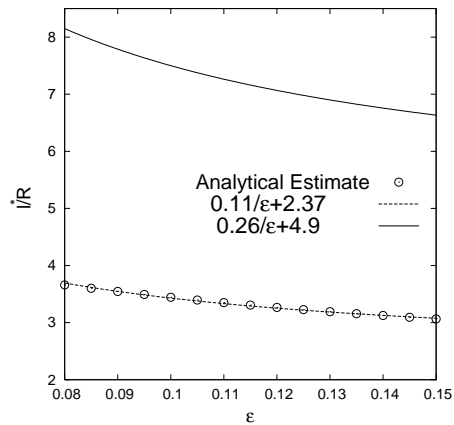}
\caption{The transport-mean-free path $l^{\ast}$ of our model wet
foam, as a function of the liquid volume fraction $\varepsilon$.
$l^{\ast}$ is measured in units of the bubble radius $R$.
$n_{in}=1.0$ and $n_{out}=1.34$ are assumed. Our {\it analytical}
result agrees well with the relation $ l^{\ast}/R \approx
{0.11}/{\varepsilon}+2.37 $. Numerical simulations of Ref.\
\cite{sun1} are well described by $ l^{\ast}/R \approx
{0.26}/{\varepsilon}+4.90.$
 }
\label{f4}
\end{figure}

For the glass disks ($n_{in}=1.5$) immersed in the air
($n_{out}=1.0$), we estimate $r\approx 0.12$. Thus for a packing
fraction $\phi = 0.64$, we find $l^{\ast} \approx 7 R $. For the
glass disks immersed in the water ($n_{out}=1.34$), we find
$r\approx 0.09$ and $l^{\ast} \approx 9 R $. Our
transport-mean-free paths are smaller than the experimental
values\ \cite{granular1, granular1c} by a factor of about $2$. Now
we consider a simple model for wet foams. For the air bubbles
($n_{in}=1$) immersed in the water ($n_{out}=1.34$), we estimate
$r \approx 0.20 $. Figure~\ref{f4} delineates $l^{\ast}$ (in units
of $R$) as a function of the liquid volume fraction $\varepsilon =
1-\phi$. From Fig.~\ref{f4} we find that in the range $0.08 <
\varepsilon<0.15 $, our {analytical} result agrees well with the
relation
\begin{equation}
 l^{\ast} \approx  R(\frac{0.11}{\varepsilon}+2.37 ). \label{ourempi}
\end{equation}
Using the hybrid lattice gas model for two-dimensional foams and
Fresnel's intensity reflectance, Sun and Hutzler performed
numerical simulation of photon transport\ \cite{sun1}. For $
0.02<\varepsilon<0.16$, their numerical results can be fitted to $
l^{\ast} \approx R({0.26}/{\varepsilon}+4.90 )$. Fig.~\ref{f4}
compares our analytic prediction with the numerical result of
Ref.\ \cite{sun1}. Again our analytic estimate is smaller than the
numerical simulations by a factor of about $2$.

Quite remarkably, our {\it analytic} estimate of the
transport-mean-free path $l^{\ast}$ quoted in
Eq.~(\ref{mainresult}), sheds some light on the empirical law of
Vera et al. and the numerical simulation of Sun and Hutzler:
$l^{\ast}/R$ is a {\it linear} function of $1/\varepsilon$, see
Eqs.~(\ref{empi}) and (\ref{ourempi}). For a better understanding
of the empirical law (\ref{empi}), we aim at a more realistic
model which not only considers Fresnel's intensity reflectance
with its significant dependence on the incidence angle, but also
the broad distribution of photons' step length. Also an extension
to the three-dimensional packing of polydisperse spheres is
envisaged.

%The transport-mean-free path $l^{\ast}$

\begin{acknowledgments}
We thank Iran's Ministry of Science, Research and Technology for
support of the parallel computing facilities at IASBS under Grant
No. 1026B (503495). MF.M. and Z.S. appreciate Iranian
Telecommunication Research Center (ITRC) for financial support.
\end{acknowledgments}

\appendix
\section{The probability distributions $F(\gamma) $ and $F'(\gamma')$ }\label{app1}

To find the probability distribution of the random variable
$\gamma $, we assume that the impact parameter $s$ in
Fig.~\ref{f1}(c) has a uniform distribution in the interval
$[0,R]$. In other words, we assume that the number of incident
rays with an impact parameter less than $s$ is {\it proportional}
to $s$. The cumulative distribution function $F_c(\gamma)\equiv
\int_{0}^{\gamma} F(\psi) d\psi $ is then $
F_c(\gamma)=\text{Prob}(s < R \sin\gamma)=\sin \gamma $. It
follows that
\begin{equation}
F(\gamma)=\frac{d F_c(\gamma)}{d \gamma}=\cos \gamma. \label{A1}
\end{equation}

Now we consider path of photons inside the disk, see
Fig.~\ref{f1}(d). Each ray can be characterized by its distance
$s'$ from the center of the disk. We assume that the random
variable $s'$ has a uniform distribution in the interval $[0,R]$.
Since $s'=R \sin\gamma'$, the cumulative distribution function is
${F'}_c(\gamma')=\sin\gamma'$, and
\begin{equation}
F'(\gamma')=\frac{d {F'}_c(\gamma')}{d \gamma'}=\cos \gamma'.
\label{A1}
\end{equation}
Further numerical simulations of our model confirm these
analytical results for $F(\gamma)$ and  $F'(\gamma')$.


\begin{thebibliography}{99}
\bibitem{comment}
Note the other possibility: A photon moving in the host medium
along the direction $\theta+\pi-2\gamma $ may be reflected into
the direction $ \theta$. However, $
\int_{0}^{{\pi}/{2}}P_{n}(x-\bar{L}\cos\theta, y-\bar{L}\sin\theta
| \theta+\pi-2\gamma)~   F(\gamma) d\gamma=\int_{-{\pi}/{2}}^{0}
P_{n}(x-\bar{L}\cos\theta, y-\bar{L}\sin\theta |
\theta+\pi+2\gamma)~F(\gamma) d\gamma $. The factor $1/2$ in the
first term of Eq.~(\ref{master}) accounts for the fact that
movement in the directions $\theta+\pi \pm 2\gamma$ have the same
probability.
\end{thebibliography}
\end{document}